\documentclass[a4paper,11pt]{article}
\pdfoutput=1 

\usepackage{jheppub} 
\usepackage{amssymb,amsfonts,amsmath,bm,bbm}
\usepackage{graphicx}
\newcommand{\order}[1]{\mathcal{O}\left(#1\right)}  

\usepackage[T1]{fontenc} 

\title{\boldmath QCD $\theta$-vacuum energy and axion properties}

\author[a,b]{Zhen-Yan Lu,}\emailAdd{luzhenyan@hnust.edu.cn}
\author[c]{Meng-Lin~Du,}\emailAdd{du@hiskp.uni-bonn.de}
\author[b,d]{Feng-Kun~Guo,}\emailAdd{fkguo@itp.ac.cn}
\author[c,e,f]{Ulf-G.~Mei\ss{}ner,}\emailAdd{meissner@hiskp.uni-bonn.de}
\author[c]{Thomas~Vonk}\emailAdd{vonk@hiskp.uni-bonn.de}

\affiliation[a]{School of Physics and Electronic Science, Hunan University of Science and Technology, \\Xiangtan 411201, China}

\affiliation[b]{CAS Key Laboratory of Theoretical Physics, Institute of Theoretical Physics, 
              \\Chinese Academy of Sciences, Beijing 100190, China}
            
\affiliation[c]{Helmholtz-Institut f\"ur Strahlen- und
             Kernphysik and Bethe Center for Theoretical Physics,\\
             Universit\"at Bonn,  D-53115 Bonn, Germany}
             
\affiliation[d]{School of Physical Sciences,
            University of Chinese Academy of Sciences,\\
            Beijing 100049, China}
             
\affiliation[e]{\mbox{Institute for Advanced Simulation, Institut f\"ur Kernphysik and J\"ulich Center for Hadron Physics,}
\mbox{Forschungszentrum J\"ulich, D-52425 J\"ulich, Germany}}

\affiliation[f]{Tbilisi State University, 0186 Tbilisi, Georgia}

\abstract{
At low energies, the strong interaction is governed by the Goldstone bosons associated with the spontaneous chiral symmetry breaking, which can be systematically described by chiral perturbation theory.
In this paper, we apply this theory to study the $\theta$-vacuum energy density and hence the QCD axion potential up to
next-to-leading order with $N$ non-degenerate quark masses.
By setting $N=3$, we then derive the axion mass, self-coupling, topological susceptibility and the normalized fourth cumulant both analytically and numerically, taking the strong isospin breaking effects into account.
In addition, the model-independent part of the axion-photon coupling, which is important for axion search experiments, is also extracted from the chiral Lagrangian supplemented with the anomalous terms up to $\mathcal{O}(p^6)$.

}

\begin{document} 
\maketitle
\flushbottom


\section{Introduction}

A CP-violating topological term, i.e., the $\theta$-term,
is allowed in the Quantum Chromodynamics (QCD) Lagrangian.
It  can be written as
\begin{eqnarray}
\mathcal{L}_\theta= \theta_0 \frac{\alpha_s}{8\pi} G^{\mu\nu,c}\tilde{G}_{\mu\nu}^c,
\label{eq:Ltheta}
\end{eqnarray}
where $\alpha_s$ is the QCD coupling constant, $G^{\mu\nu,c}$ is the gluon field strength tensor, with $c$ a color index, and
$\tilde{G}_{\mu\nu}^c=\varepsilon_{\mu\nu\rho\sigma}G^{\rho\sigma,c}/2$  its dual.
Because none of the quarks is massless, physical observables only depend on a combination of the $\theta_0$ parameter  and the phases present in the quark mass matrix $\mathcal{M}_q$, i.e., $\theta=\theta_0+\mathrm{arg~det}\mathcal{M}_q$.
Being a dimensionless parameter, the natural value of $\theta$ is expected to be $\mathcal{O}(1)$, which would  significantly affect physical systems such as atomic nuclei, and lead to measurable effects, as nucleons, for instance, would possess a nonvanishing electric dipole moment~\cite{79Crewther.DiVecchia.ea123-123PLB}.
However, the so-far negative results of experimental searches for the nucleon electric dipole moment lead to a tiny upper limit: $|\theta|\lesssim10^{-10}$~\cite{06Baker.others131801-131801PRL,09Griffith.Swallows.ea101601-101601PRL,15Parker.others233002-233002PRL,16Graner.Chen.ea161601-161601PRL,15Guo.Horsley.ea62001-62001PRL,Abel:2020gbr}. To understand why the value of $\theta$ is so small is the so-called {\em strong CP problem}.
One elegant possible solution of this problem is the Peccei--Quinn (PQ) mechanism~\cite{77Peccei.Quinn1440-1443PRL,77Peccei.Quinn1791-1797PRD}, which introduces a global U(1) symmetry, called PQ symmetry. This symmetry is spontaneously broken at energies much higher than the typical QCD scale  of order
$\mathcal{O}(1\,\text{GeV})$ and is also broken by an anomalous coupling to gluon fields.
The axion appears as the corresponding Goldstone boson~\cite{78Weinberg223-226PRL,78Wilczek279-282PRL} which has an anomalous coupling to $G\tilde{G}$. The parameter $\theta$ is then dynamically driven to zero at the minimum of the axion potential, giving rise to a possible solution to the strong CP problem.

In the past few decades, there have been tremendous efforts  searching for the axion, denoted by $a$, as well as constraining to its mass $m_a$ and decay constant $f_a$, see, e.g., \cite{09Arik.others8-8J,08Inoue.Akimoto.ea93-97PLB,15Arik.others21101-21101PRD,10Asztalos.others41301-41301PRL,13Armengaud.others67-67J,13Keller.Sedrakian62-69NPA,
15Arvanitaki.Baryakhtar.ea84011-84011PRD,17Brubaker.others61302-61302PRL,17Fu.others181806-181806PRL,17Liu.others52006-52006PRD,10Ehret.others149-155PLB,15Ballou.others92002-92002PRD,03Bradley.Clarke.ea777-817RMP}.
Some important quantities in  axion physics, such as the axion mass and self-coupling, are dictated by the axion potential.
The visible axion models~\cite{78Weinberg223-226PRL,78Wilczek279-282PRL} with the axion decay constant at the electroweak scale or even smaller are believed to have been ruled out by experiments.
For the invisible axion~\cite{81Dine.Fischler.ea199-202PLB,80Zhitnitsky260-260SJNP,79Kim103-103PRL,Shifman-1979if}, its mass window is usually assumed in the range from about $10^{-6}$~eV to $10^{-2}$~eV.
According to constraints from astrophysical observations, the present bounds on the axion decay constant is
$
10^9~\mathrm{GeV}\lesssim f_a\lesssim10^{12}~\mathrm{GeV}
$
~\cite{87Kim1-177PR,88Raffelt.Seckel1793-1793PRL} (we refer to Refs.~\cite{03Bradley.Clarke.ea777-817RMP,10Kim.Carosi557-602RMP,16Marsh1-79PR,Tanabashi:2018oca,Luzio:2020} for several recent reviews).\footnote{It was recently argued that there is still a possibility for a viable QCD axion model with a mass in the MeV range~\cite{18Alves.Weiner92-92J}.} Within the above available parameter space,
the axion may be the main source of cold dark matter in the universe~\cite{83Preskill.Wise.ea127-132PLB,Abbott-1982af,83DinePLB,09Visinelli.Gondolo35024-35024PRD,14Visinelli.Gondolo11802-11802PRL,17Kaneta.Lee.ea101802-101802PRL}.
In addition, it may form a Bose--Einstein condensate~\cite{09Sikivie.Yang111301-111301PRL} or even compact boson stars~\cite{93Kolb.Tkachev3051-3054PRL,16Bai.Barger.ea127-127J,16Braaten.Mohapatra.ea121801-121801PRL,16Eby.Leembruggen.ea66-66J,17Levkov.Panin.ea11301-11301PRL,18Visinelli.Baum.ea64-72PLB,Braaten-2019knj}.
The axion can couple to the Standard Model (SM) particles like electrons, nucleons, photons and so on. However,
all these couplings are suppressed by the axion decay constant $f_a$, which is remarkably large, resulting in the invisible axion which has very weak couplings to the SM particles~\cite{83Preskill.Wise.ea127-132PLB}.
Since the axion-photon coupling vertex, see Eq.~(\ref{garr-coupling}) below, allows for the production of an axion from the interaction of a photon with a background magnetic field, the axion-photon coupling $g_{a\gamma\gamma}$
plays a central role in axion searches in both laboratory experiments and stellar objects~\cite{03Bradley.Clarke.ea777-817RMP}.
In this case it is very useful  to study the axion properties, especially the axion-photon coupling, at a high precision
from the theoretical point of view.

At low energies in QCD, all hadronic degrees of freedom are frozen and thus can be neglected except for the pseudo-Goldstone bosons of the spontaneous chiral symmetry breaking. Chiral perturbation theory (CHPT)~\cite{79Weinberg327-340PA,84Gasser.Leutwyler142-142AP,85Gasser.Leutwyler465-516NPB}, as the low-energy effective theory of QCD,
can be used to describe the vacuum properties as well as the dynamics of QCD in the non-perturbative regime reliably.
In this paper, we will calculate the $\theta$-vacuum energy density, or equivalently the QCD axion potential,
up-to-and-including  next-to-leading order (NLO) in SU($N$) CHPT.
Setting $N=3$, the mass and self-couplings of the axion can then be extracted from a Taylor expansion of the axion potential.
In addition, we also compute the NLO corrections to the axion-photon coupling.

Before continuing, we would like to stress that a similar study was performed in Ref.~\cite{16GrillidiCortona.Hardy.ea34-34J}, where the QCD axion potential derived in two-flavor CHPT up to NLO (the QCD $\theta$-vacuum energy density up to NLO was first derived in Ref.~\cite{15Guo.Meisner278-282PLB}) is used, and a matching between two-flavor and three-flavor CHPT is performed to determine the axion-photon coupling. Here, the calculations are explicitly done in SU($N$) CHPT for the $\theta$-vacuum energy density and with $N=3$ for the other quantities.
In SU(3) CHPT, the topological susceptibility as well as the fourth cumulant of the topological distribution up to NLO have been calculated before using the Goldstone boson masses at $\theta=0$~\cite{09Mao.Chiu34502-34502PRD,12Bernard.Descotes-Genon.ea80-80J,18Luciano.Meggiolaro74001-74001PRD}.
Very recently, the topological susceptibility and axion mass are calculated up to next-to-next-to-leading order and including electromagnetic corrections up to $\order{\alpha_\text{em}}$ in SU(2) CHPT ~\cite{19Gorghetto.Villadoro33-33J}.
The axion-nucleon coupling is also calculated up to the leading one-loop order in Ref.~\cite{Vonk:2020zfh}.
Here, we derive the one-loop contribution to the SU($N$) $\theta$-vacuum energy density by a direct calculation of the logarithm of the functional determinant for the Goldstone bosons in a $\theta$-vacuum, extending the two-flavor treatment in Ref.~\cite{15Guo.Meisner278-282PLB} to the case of $N$ non-degenerate flavors.\footnote{For an investigation of the axion interactions with mesons and photons using a 3-flavor chiral Lagrangian including the U(1)$_A$ anomaly, see Ref.~\cite{Landini:2019eck}. }
This study is useful when the up and down quark masses take values close to the strange quark one. This could happen in lattice QCD calculations where the quark masses are parameters that can be chosen freely.

The outline of the paper is as follows. In Sec.~\ref{EvacLONLO}, we generalize the calculation of the $\theta$-vacuum energy density in the framework of SU(2) CHPT~\cite{15Guo.Meisner278-282PLB} to the SU($N$) case with $N$ non-degenerate quark masses.
In Sec.~\ref{AxionPro}, we derive the axion properties, including the mass and self-coupling, in detail for $N=3$.
In Sec.~\ref{garr-su3}, the model-independent part of the axion-photon coupling is determined from the chiral Lagrangian supplemented with the odd-intrinsic-parity sector of the chiral effective Lagrangian. Section~\ref{SUMMARY} contains a brief summary and discussion. The appendix provides a relatively detailed derivation of the recursion relation giving rise to the
general solution of the vacuum angles $\phi_f$.

\section{\texorpdfstring{$\bm{\theta}$}{Theta}-vacuum energy density up to NLO}
\label{EvacLONLO}

The QCD axion potential as a function of $a/f_a$ has the same form as the QCD $\theta$-vacuum energy density as a function of $\theta$. In this section, we compute the $\theta$-vacuum energy density in SU($N$) CHPT with $N$ non-degenerate quark masses, which is an extension of Ref.~\cite{15Guo.Meisner278-282PLB}, where the $\theta$-vacuum energy is computed up to NLO in the SU(2) and SU($N$)-symmetric cases.

\subsection{Leading order}

The discovery of instantons not only solved the $U(1)_A$ problem, but also implied that there is a $\theta$-term in the QCD Lagrangian. In order to study the physics with a $\theta$ parameter, it is common to rotate away the $\theta$-term by performing a chiral rotation on the quark fields. At low energies, we can then match the resulting Lagrangian to the chiral Lagrangian since now the relevant degrees of freedom are the pseudo-Goldstone bosons~\cite{84Gasser.Leutwyler142-142AP,85Gasser.Leutwyler465-516NPB}.
The Lagrangian density of the SU($N$) CHPT at leading order (LO) in a $\theta$-vacuum is
\begin{align}\label{Lp2}
\mathcal{L}^{(2)}=\frac{F_0^2}{4}\Big[\langle D_\mu U D^\mu U^{\dagger}\rangle+\langle\chi_\theta^{} U^{\dagger} +U\chi_\theta^{\dagger}\rangle\Big],
\end{align}
where  $\chi_\theta^{}=2B_0\mathcal{M}_q\exp[-i\mathcal{X}_a\theta] $ contains the $\theta$ angle (in the convention that $U\to g_R^{}Ug_L^\dagger$ under chiral transformations, see section~\ref{garr-su3}) and  the diagonal and real quark mass matrix is $\mathcal{M}_q=\mathrm{diag}\{m_1,m_2,...,m_N\}$, and $\langle\ldots\rangle$ denotes the trace in the flavor space. The matrix $\mathcal{X}_a$ takes the following general form:
\begin{eqnarray}\label{Xa}
\mathcal{X}_a=\text{diag}\{\mathcal{X}_1,\mathcal{X}_2,...,\mathcal{X}_N\},\quad \langle\mathcal{X}_a \rangle=1,
\end{eqnarray}
which arises from a U(1)$_A$ chiral rotation on the quark fields eliminating the $\theta$-term in the QCD Lagrangian. In this case, the $\theta$-dependence is completely captured by the quark mass term.
The U(1)$_A$ chiral rotation can be distributed to different quark flavors, leading to different choices of $\mathcal{X}_a$.
$F_0$ is the pion decay constant in the three-flavor chiral limit, and $B_0=-\langle\bar q q\rangle/F_0^2$ is related to the scalar quark condensate. $U(x)$ is the field configuration for the vacuum and the Goldstone bosons of the spontaneous breaking of chiral symmetry. It can be written as  $U(x)=U_0\tilde{U}(x)$, where $\tilde{U}(x)$ collects the Goldstone bosons, and $U_0$ describes the vacuum, parameterized as
\begin{eqnarray}
U_0=\mathrm{diag}\{e^{i\varphi_1},e^{i\varphi_2},...,e^{i\varphi_N}\}
\label{eq:U0}
\end{eqnarray}
subject to the  constraint $\sum_{i=1}^N\varphi_i=0$~\cite{03Brower.Chandrasekharan.ea64-74PLB,09Mao.Chiu34502-34502PRD}.
For the SU(3) case, $\tilde{U}=e^{i\Phi/F_0}$, with $\Phi$ given by
\begin{eqnarray}
\Phi=\left(
            \begin{array}{ccc}
              {\pi_3+\frac{1}{\sqrt{3}}\eta_8} & \sqrt{2}\pi^{+} & \sqrt{2}K^{+} \\
              \sqrt{2}\pi^{-} & {-\pi_3+\frac{1}{\sqrt{3}}\eta_8} & \sqrt{2}K^0 \\
              \sqrt{2}K^{-} & \sqrt{2}\bar{K}^0 & -{\frac{2}{\sqrt{3}}\eta_8} \\
            \end{array}
          \right).
\end{eqnarray}
Note that the neutral flavor eigenstates in the octet of the pseudoscalar mesons as shown above, i.e. $\pi_3$ and $\eta_8$, are not mass eigenstates. Diagonalizing the mass matrix of the meson fields, one gets the physical mass eigenstates $\pi^0$ and $\eta$, which are mixtures of  $\pi_3$ and $\eta_8$.
By expanding the LO Lagrangian in terms of the meson fields to quadratic order, the LO $\theta$-dependent meson masses including isospin breaking effects are obtained as
\begin{eqnarray}
\mathring{M}_{\pi^{\pm}}^2&=& B_0(m_1\cos\phi_1+m_2\cos\phi_2), \nonumber\\
\quad \mathring{M}_{K^{\pm}}^2&=&B_0 (m_1\cos\phi_1+m_3\cos\phi_3), \nonumber\\
\mathring{M}_{K^0}^2&=&\mathring{M}_{\bar{K}^0}^2=B_0(m_2\cos\phi_2+m_3\cos\phi_3), \nonumber\\
\mathring{M}_{\pi^{0}}^2&=&B_0(m_1\cos\phi_1+m_2\cos\phi_2) -\xi  ,
\nonumber\\
\mathring{M}_\eta^2&=&\frac{1}{3}B_0\left(m_1\cos\phi_1+m_2\cos\phi_2+4m_3\cos\phi_3 \right) + \xi~,
\label{M(theta)-eta}
\end{eqnarray}
where for convenience we have defined   $(m_1,m_2,m_3)\equiv(m_u,m_d,m_s)$ and
\begin{equation}
\phi_f \equiv \mathcal{X}_f\theta+\varphi_f.
\label{eq:phif}
\end{equation}
The parameter $\xi$ is given by
\begin{eqnarray}
\xi=\frac{4}{3}B_0\left(m_3\cos\phi_3 -\frac{1}{2}(m_1\cos\phi_1+m_2\cos\phi_2) \right) \frac{\sin^2\epsilon_\theta}{\cos(2\epsilon_\theta)} = \mathcal{O}\left(\epsilon_\theta^2\right),
\end{eqnarray}
with $\epsilon_\theta$ the $\pi^0$-$\eta$ mixing angle in the $\theta$-vacuum, which arises due to strong isospin breaking. Diagonalization of the mass matrix requires
\begin{eqnarray}\label{eeTheta}
\tan 2\epsilon_\theta = \frac{\sqrt{3}(m_2\cos\phi_2-m_1\cos\phi_1)}{2m_3\cos\phi_3-m_1\cos\phi_1-m_2\cos\phi_2} ~.
\end{eqnarray}
Obviously, the above $\theta$-dependent Goldstone boson masses reduce to the standard SU(3) relations~\cite{85Gasser.Leutwyler465-516NPB} by taking the limit  $\theta = 0$ and setting $\phi_f=0$.
The dependence of $\phi_f$ on the $\theta$ angle needs to be determined by minimizing the vacuum energy to be discussed below.

To determine the ground state, i.e. the vacuum, we set $\tilde{U}=\mathbbm{1}$.
Performing the trace in Eq.~(\ref{Lp2}), one obtains the LO potential energy density
\begin{eqnarray}\label{LP22}
e^{(2)}_\text{vac}= - F_0^2B_0\sum_{f}m_f\cos\phi_f .
\end{eqnarray}
Moreover, minimizing Eq.~(\ref{LP22}) with respect to the parameters $\phi_f$ with the constraint $\sum_f \phi_f = \theta$ gives
the following equations\footnote{For the vacuum alignment in SU(2) CHPT up to  NLO, we refer to the appendix of Ref.~\cite{15Acharya.Guo.ea54023-54023PRD}, which also shows that it is sufficient to consider the LO vacuum alignment for the computation of the cumulants up to $\order{p^4}$. }
\begin{eqnarray}
\begin{cases}
m_1\sin\phi_1 = m_2\sin\phi_2 = m_3\sin\phi_3 ,\cr
\sum_f \phi_f = \theta ,
\end{cases}
\label{eq:dashen-nuyts}
\end{eqnarray}
for SU(3), and similar equations for SU($N$), i.e., $m_f\sin\phi_f$ is the same for all flavors.
The above equations depend only on the linear combination $\phi_f$ given in Eq.~\eqref{eq:phif}, instead of on $\mathcal{X}_f$ and $\varphi_f$ separately.
This implies that $\phi_f$ is physical while $\mathcal{X}_f$ and $\varphi_f$ are not.
One can use this freedom to choose the  ``gauge" most convenient for the question of interest.
One possible choice is to choose $\mathcal{X}_a=\mathbbm{1}/N$, which is commonly used in the literature  (see, e.g., Refs.~\cite{03Brower.Chandrasekharan.ea64-74PLB,09Mao.Chiu34502-34502PRD,10Aoki.Fukaya34022-34022PRD,15Guo.Meisner278-282PLB,15Acharya.Guo.ea54023-54023PRD}).
Noticing that the only constraint on $\mathcal{X}_a$ is $\langle\mathcal{X}_a\rangle=1$, one may also choose the U(1)$_A$ rotations to be
\begin{equation}
\mathcal{X}_f = \frac{\phi_f}{\theta},\quad \text{and}\quad \varphi_f = 0
\label{eq:Xf}
\end{equation}
to simultaneously shift the $\theta$ angle to the quark mass matrix phase and align the vacuum properly. This is a convenient choice for the $a\gamma\gamma$ coupling (with $\theta$ changed to the dynamical axion field $a/f_a$) to be discussed in Sec.~\ref{garr-su3} since this removes the leading order $a$-$\pi^0$ and $a$-$\eta$ mixing.

The equations~\eqref{eq:dashen-nuyts} do not admit an analytical solution in terms of elementary functions in a compact form.\footnote{In the SU(2) case, there is an analytic solution~\cite{03Brower.Chandrasekharan.ea64-74PLB}, which then allows to derive a closed form of the vacuum energy density up to NLO in the chiral expansion~\cite{15Guo.Meisner278-282PLB}.}
In the isospin symmetric case,
the up and down quark masses are degenerate $m_1=m_2\equiv m$ but $ m\neq m_3$, we have
$\phi_1=\phi_2\equiv\phi$, and then Eq.~(\ref{eq:dashen-nuyts}) becomes~\cite{18Bernard.Toussaint74502-74502PRD}
\begin{eqnarray}
m\sin\phi = {m_3} \sin(\theta-2\phi) ,
\label{eq:dashen22}
\end{eqnarray}
which allows for analytic solutions, though complicated ones.

If one focuses on the cumulants of the QCD topological distribution, which are derivatives of the vacuum energy density, $e_\text{vac}(\theta)$, with respect to $\theta$,
\begin{equation}
    c_{2n} = \left.\frac{d^{2n} e_\text{vac}(\theta)}{d\theta^{2n}} \right|_{\theta = 0}, \quad n\in \mathbbm{N} ,
    \label{eq:cumulants}
\end{equation}
one may solve Eqs.~\eqref{eq:dashen-nuyts} by expanding in powers of $\theta$.
Specifically, $c_2$ corresponds to the  topological susceptibility.
Up to  $\order{\theta^3}$, one gets~\cite{09Mao.Chiu34502-34502PRD}
\begin{eqnarray}\label{phiSolu}
\phi_f&=&\frac{\bar{m}}{m_f}\theta+\bigg[\Big(\frac{\bar{m}}{m_f}\Big)^3
-\frac{\bar{m}^4}{m_f\bar{m}^{[3]}}
\bigg]\frac{\theta^3}{6}+\mathcal{O}(\theta^5),
\end{eqnarray}
where we have introduced
\begin{eqnarray}
\frac{1}{\bar{m}}=\sum_{i}\frac{1}{m_i}~,~~~	
\frac{1}{\bar{m}^{[3]}}=\sum_{i}\frac{1}{m_i^3}
\label{eq:mbar}
\end{eqnarray}
with $i$ running over all the flavor indices considered in the theory. The solutions in Eq.~(\ref{phiSolu}) are not restricted to the three-flavor case but also valid for $N>3$.
Consequently, the $\theta$-dependence of the vacuum energy density at LO
can be obtained by substituting the solution in Eq.~\eqref{phiSolu} into Eq.~\eqref{LP22}, which gives~\cite{09Mao.Chiu34502-34502PRD}
\begin{eqnarray}\label{VaLO}
e_{\text{vac}}^{(2)}(\theta)
=F_0^2B_0\bigg(\frac{1}{2}\bar{m}\theta^2
-\frac{\bar{m}^4}{24\bar{m}^{[3]}}\theta^4 \bigg)+\mathcal{O}(\theta^6).
\end{eqnarray}
In App.~\ref{app:general}, we work out a recursion relation for $\phi_f$ up to an arbitrary power of $\theta$,
\begin{eqnarray}\label{phifSolugener}
\phi_f &=&  \sum_{n=0}^{\infty} C_{f,2n+1} \theta^{2n+1},
\end{eqnarray}
with $C_{f,1}=\bar m/m_f$ and
\begin{eqnarray}
 C_{f,2n+1} &=& \sum_{t=1}^{n} \sum_{(k_1,\ldots,k_t)}  s_{K_t}^{}\binom{K_t}{k_1,\ldots,k_{t}} \Bigg[ \frac{\bar m}{m_f} \sum_{i=1}^N \prod_{j=1}^{t} C_{i,2j-1}^{k_j}
 - \prod_{j=1}^{t} C_{f,2j-1}^{k_j}  \Bigg],~~~~ \label{eq:recursion}
\end{eqnarray}
where $k_j$ are non-negative integers, $K_t \equiv\sum_{j=1}^t k_j$, $\sum_{(k_1,\ldots,k_t)}$ means that the sum runs over all possibilities of $k_j$ satisfying $k_1+\cdots+(2t-1)k_t=2n+1$, $s_{K_t}^{} = (-1)^{(K_t-1)/2}/(K_t!)$, and $\binom{K_t}{k_1,\ldots,k_{t}}= K_t!/(k_1!\cdots k_t!)$ are the multinomial coefficients.

In the next subsection, we will compute the one-loop contribution of the Goldstone bosons to the energy density.

\subsection{Next-to-leading order}

To study the $\theta$-vacuum energy up to the NLO, $\mathcal{O}(p^4)$, one has to include both the tree-level diagrams from
$\mathcal{L}^{(4)}$ and the one-loop diagrams with innsertions from
 $\mathcal{L}^{(2)}$. The SU($N$) chiral Lagrangian at NLO is given by
\begin{eqnarray}\label{LOp4}
\mathcal{L}^{(4)} &=&
L_6\langle\chi_\theta U^{\dagger} +U\chi_\theta^{\dagger}\rangle^2 +L_7\langle\chi_\theta^{} U^{\dagger} -U\chi_\theta^{\dagger}\rangle^2 \nonumber\\ &&+L_8\langle\chi_\theta^{\dagger}U\chi_\theta^{\dagger}U+U^{\dagger}\chi_\theta^{} U^{\dagger}\chi_\theta^{}  \rangle +H_2\langle\chi_\theta^{\dagger}\chi_\theta^{}\rangle  ,~~
\end{eqnarray}
where we only display the terms relevant for the vacuum energy. The $L_i$ and $H_2$ are the so-called  low-energy constants (LECs) and the high-energy constant (HEC), respectively. The latter is only required for renormalization and does not appear in observables.
After setting $U=U_0$ and evaluating the traces, one gets the tree-level contribution to the NLO vacuum energy density
\begin{eqnarray}
e^{(4,\mathrm{tree})}_{\mathrm{vac}} &=&
- 16B_0^2 \bigg[L_6\bigg(\sum_i m_i\cos\phi_i\bigg)^2
- L_7\bigg(\sum_im_i\sin\phi_i\bigg)^2 \nonumber\\
&&+\frac{L_8}{2}\sum_im_i^2\cos(2\phi_i)
+\frac{H_2}{4}\sum_im_i^2 \bigg].
\label{eq:etree}
\end{eqnarray}
The LECs and HEC contain both ultraviolet (UV) finite and divergent parts. They are related to the renormalized ones, denoted by an upper index $r$, by~\cite{85Gasser.Leutwyler465-516NPB,09Bijnens.Lu116-116J}
\begin{eqnarray}\label{LECSU3}
L_6=L_6^r+\frac{N^2+2}{16N^2}\lambda,\quad L_8=L_8^r+\frac{N^2-4}{16 N}\lambda,\quad L_7=L_7^r,\quad H_2=H_2^r+\frac{N^2-4}{8N}\lambda ,~~~~
\end{eqnarray}
with
\begin{equation}
 \lambda = \frac{\mu^{d-4}}{16\pi^2}\left\{\frac1{d-4} - \frac12 \left[\ln(4\pi) +\Gamma'(1) +1 \right] \right\}
\end{equation}
the UV divergence at the space-time dimension $d=4$, where $\mu$ is the scale of  dimensional regularization.
The UV divergence in the NLO tree-level contribution exactly cancels the one arising in the one-loop contribution, as will be seen below.

Now let us calculate the one-loop contribution to the $\theta$-vacuum energy density.
In the classical  CHPT papers~\cite{84Gasser.Leutwyler142-142AP,85Gasser.Leutwyler465-516NPB},
the one-loop effective generating functional is expanded around the free-field configuration at $\theta=0$. This treatment is then applied to derive the topological susceptibility and the fourth cumulant in SU($N$) CHPT in Refs.~\cite{09Mao.Chiu34502-34502PRD,10Aoki.Fukaya34022-34022PRD,12Bernard.Descotes-Genon.ea80-80J}. The expression for the vacuum energy density at  NLO in SU(2) with non-degenerate quark masses, as well as that in SU($N$) with degenerate quark masses, is derived in Ref.~\cite{15Guo.Meisner278-282PLB}, where the generating functional is expanded around the free-field configuration in the $\theta$-vacuum. The result allows for an evaluation of any cumulant of the QCD topological charge distribution, and is the QCD axion potential at NLO~\cite{16GrillidiCortona.Hardy.ea34-34J}.
Here, we generalize the result in Ref.~\cite{15Guo.Meisner278-282PLB} to SU($N$), with $N$ non-degenerate quark masses. The effective action for the free-field configuration in the $\theta$-vacuum is
\begin{eqnarray}
Z_0(\theta)&=&\frac{i}{2}\ln\det D_0(\theta)=\frac{i}{2}\mathrm{Tr}\ln D_0(\theta),
\end{eqnarray}
where ``Tr" denotes traces over both the flavor (in the adjoint representation) and the coordinate spaces,
and the differential operator $D_0(\theta)$ takes the following form
\begin{eqnarray}
D_{0,PY}(\theta)&=& \delta_{PY} \left[\partial^\mu\partial_\mu + \mathring{M}_P^2(\theta) \right],\label{DDDa}
\end{eqnarray}
where $P,Y=1,\ldots,N^2-1$ are the flavor indices of the Goldstone bosons, and $\mathring{M}_P(\theta)$ are $\theta$-dependent meson masses at LO given in Eq.~\eqref{M(theta)-eta}.
Within dimensional regularization, one gets the one-loop contribution to the vacuum energy density as~\cite{15Guo.Meisner278-282PLB}
\begin{eqnarray}\label{L-loop}
e_{\text{vac}}^{\mathrm{(4,loop)}}&=& -\frac{Z_0(\theta)}{V}  \nonumber\\
&=& - \frac{i}{2} \sum_P \int \frac{d^dp}{(2\pi)^d} \ln \left[-p^2 +\mathring{M}_P^2(\theta) \right] \nonumber\\
&=&\sum_P \mathring{M}_P^4(\theta)\left\{\frac{\lambda}{2} - \frac{1}{128\pi^2}\left[
1-2\ln\frac{\mathring{M}_P^2(\theta)}{\mu^2}\right]\right\},
\label{eq:eloop}
\end{eqnarray}
where $V$ is the space-time volume, the $P$ runs over the Goldstone boson mass eigenstates (for the SU(3) case, they are given in Eq.~\eqref{M(theta)-eta}),
and the  term proportional to $\lambda$ collects all the UV divergences in the one-loop contribution.

Noticing that the matrix elements of the diagonalized mass-squared matrix of the Goldstone bosons are given by
\begin{eqnarray}
 \delta_{PY}\mathring{M}_{P}^2(\theta) &=& \frac{1}{8}\left\langle \left\{\lambda_P,\lambda_Y^\dagger\right\}\left(\chi_\theta^\dagger U_0+U_0^\dagger \chi_\theta^{}\right)\right\rangle
 \equiv \sigma_{PY},
\end{eqnarray}
we obtain
\begin{eqnarray}
\sum_P\mathring{M}_{P}^2(\theta)
&=&\sum_P\sigma_{PP}^{} =
\frac{2(N^2-1)}{N}\sum_i m_i\cos\phi_i.
\end{eqnarray}
Similarly, we have
\begin{eqnarray}
\sum_P\mathring{M}_{P}^4(\theta) &=&\sum_{P,Y} \sigma_{PY}^{} \sigma_{YP}^{} \nonumber\\
&=& 2B_0^2\bigg[\frac{N^2+2}{N^2}\Big(\sum_i m_i\cos\phi_i \Big)^2
+\frac{N^2-4}{N}\sum_i (m_i\cos\phi_i)^2 \bigg].
\label{eq:mp4}
\end{eqnarray}
With Eqs.~\eqref{eq:etree}, \eqref{LECSU3}, \eqref{eq:eloop} and \eqref{eq:mp4}, it is straightforward to check that the UV divergence in the one-loop contribution exactly cancels that in the tree-level contribution. Finally, we obtain the $\theta$-vacuum energy density up to NLO as
\begin{eqnarray}
 e_\text{vac} &=&
  - F_0^2 B_0 \sum_i m_i\cos\phi_i - \sum_P \frac{\mathring{M}_P^4(\theta)}{128\pi^2} \left[1-2\ln\frac{\mathring{M}_P^2(\theta)}{\mu^2}\right] \nonumber\\
   && -16 B_0^2 \Bigg[L_6^r \Big(\sum_i m_i\cos\phi_i\Big)^2 + N \left(N L_7^r +L_8^r \right) m_1^2\cos^2\phi_1  \Bigg]
  ,
  \label{VaNLO}
\end{eqnarray}
where we have used the SU($N$) version of Eq.~\eqref{eq:dashen-nuyts} to replace all $m_i\sin\phi_i$ by $m_1\sin\phi_1$, and have neglected the $\theta$-independent terms.

From the above $\theta$-vacuum energy density, the lowest two cumulants of the topological charge distribution up to NLO can
then be easily extracted. It can be checked from Eq.~(\ref{VaNLO}) that we can reproduce the expression of topological susceptibility at NLO keeping all orders in strong isospin breaking exactly given in Ref.~\cite{12Bernard.Descotes-Genon.ea51-51J}.
We are more interested in the axion mass and its self-coupling, and thus we will extract them from the axion potential based on the relation between the $\theta$-vacuum energy and axion potential
in the following section. Numerical values of the topological susceptibility and the normalized fourth cumulant will also be given for reference.

\section{Axion mass and self-coupling}\label{AxionPro}

Both the axion mass and self-coupling are important quantities, since they directly affect experimental searches for the axion.
For example, one tries to detect the axion in microwave cavities by stimulating their conversion to photons via the Primakoff effect within an external magnetic field~\cite{03Bradley.Clarke.ea777-817RMP}.
The axion self-coupling  plays an important role in the formation of an axion Bose-Einstein condensation~\cite{09Sikivie.Yang111301-111301PRL} as well as  possible boson stars~\cite{93Kolb.Tkachev3051-3054PRL,16Braaten.Mohapatra.ea121801-121801PRL,16Bai.Barger.ea127-127J,17Levkov.Panin.ea11301-11301PRL,17Eby.Leembruggen.ea14-14J}. This motivates the study of these two quantities in this section to high precision. Before we proceed to derive the axion mass and self-coupling up to NLO,
let us discuss a little bit about the axion solution to the strong CP problem, and start with the effective Lagrangian,
\begin{eqnarray}\label{eq:gaa}
\mathcal{L}_{G\tilde{G}}=
\left(\theta+\frac{a}{f_a}\right)\frac{g_s^2}{32\pi^2}G^c_{\mu\nu}\tilde{G}^{c,\mu\nu}~,
\end{eqnarray}
where in addition to the $\theta$-term, a pseudoscalar axion field is introduced which couples to gluons. As shown by Peccei and Quinn~\cite{77Peccei.Quinn1440-1443PRL,77Peccei.Quinn1791-1797PRD}, the periodicity of the vacuum expectation value~(VEV) $\langle G\tilde G\rangle$  in $\theta+a/f_a$ forces the minimum of the axion VEV to be at $\theta + \langle a\rangle/f_a = 0$, and thus the $\theta$-dependence is eliminated. Expanding the axion field around its VEV, one sees that the
$\theta$-vacuum energy density derived in the previous section,  with $\theta$ being replaced by $a_\text{phys}/f_a$, gives the axion potential, where $a_\text{\rm phys}=a-\langle a\rangle$ is the physical axion field. In the following we will denote $a_\text{\rm phys}$ as $a$ for simplicity, and then the axion potential is given by  $V(a)=e_\text{vac}(a/f_a)$.

Expanding $V(a)$ in powers of the axion field around the vacuum, we obtain
\begin{eqnarray}\label{Va-expanding}
V(a)=\frac{1}{2}m_a^2a^2 +\sum_{n=2}^{\infty}\frac{1}{(2n)!}\lambda_{2n}a^{2n} ~.
\end{eqnarray}
Comparing the above equation with the definition of cumulants of the QCD topological distribution in Eq.~(\ref{eq:cumulants}),  one finds the following relations for the axion mass and axion self couplings:
\begin{equation} \label{malachic4}
    m_a^2=\frac{c_2}{f_a^2},\quad  \lambda_{2n}=\frac{c_{2n}}{f_a^{2n}} ~,
\end{equation}
where $c_{2n}$ are the cumulants defined in Eq.~\eqref{eq:cumulants} with $n\geq2$.
Thus, the axion mass and four-axion self-coupling at LO are given by
\begin{eqnarray}
m_{a,\text{LO}}^2 = \frac{F_\pi^2M_{\pi^+}^2\bar{m}}{2f_a^2\hat m}, \quad
\lambda_{4,\text{LO}} = -\frac{F_\pi^2M_{\pi^+}^2\bar{m}^4}{2f_a^4 \bar{m}^{[3]} \hat m} ,
\end{eqnarray}
respectively, where $\hat m=(m_u+m_d)/2$, and we have replaced $B_0$ and $F_0$ by $M_{\pi^{+}}^2/(2\hat m)$ and $F_\pi$, the physical pion mass squared and decay constant, respectively, which is legitimate at LO.
One sees that at LO, the difference between the SU(3) and SU(2) expressions resides merely in the definitions of $\bar m$ and $\bar m^{[3]}$ in Eq.~\eqref{eq:mbar}.

In the same way we have calculated the axion mass and self-couplings at LO. Their expressions at NLO, including the higher order corrections, can be extracted from Eq.~\eqref{VaNLO}. The former reads
\begin{eqnarray}
m_a^2
&=&\frac{F_\pi^2M_{\pi^{+}}^2\bar{m}}{2f_a^2\hat m}\Bigg\{1+\frac{16M_{\pi^+}^2}{F_\pi^2}
\left[\frac{3\bar m}{\hat m}\left(3L_7^r+L_8^r\right) -L_8^r \right] \nonumber\\&& +\frac{\bar m}{m_s}\left(\mu_{\pi^0}^{}+2\mu_{\pi^+}^{}-\mu_{\eta}^{} \right) + \left( 2\frac{\bar m}{m_d}-1 \right)\mu_{K^+}^{} + \left( 2\frac{\bar m}{m_u}-1 \right)\mu_{K^0}^{}
+\mathcal{O}\left(\frac{\delta^2}{m_s^2}\right)\Bigg\} ,~~~~~
\label{eq:masq}
\end{eqnarray}
with $\mu_P^{}=\frac{M_P^2}{32\pi^2F_\pi^2}\ln\frac{M_P^2}{\mu^2}$ and $\delta=m_d-m_u$, where we have used the NLO expressions for the pion mass and decay constant~\cite{85Gasser.Leutwyler465-516NPB}:
\begin{align}
    M_{\pi^+}^2 &= B_0(m_u+m_d) \left\{ 1+ \mu_{\pi^0}^{}-\frac13\mu_\eta^{} +\frac{16B_0}{F_0^2}\left[\hat m \left(2L_8^r-L_5^r\right) + (2\hat m+m_s)\left(2L_6^r-L_4^r\right) \right] \right\}, \nonumber\\
    F_{\pi} &= F_0\left\{1 - \mu_{\pi^+}^{} -\mu_{\pi^0}^{} - \frac{\mu_{K^+}^{}}{2} -\frac{\mu_{K^0}^{}}{2} + \frac{8B_0}{F_0^2}\left[ \hat m L_5^r + (2\hat m+m_s)L_4^r \right] \right\}.
    \label{eq:mpifpi}
\end{align}
Similarly the self-coupling up to NLO can be easily obtained as
\begin{align}
\lambda_4
=& - \frac{F_\pi^2M_{\pi^{+}}^2\bar{m}^4}{2f_a^4\hat{m}\bar{m}^{[3]}} \Bigg\{ 1
+ \frac{16M_{\pi^+}^2 }{F_\pi^2}\left[\frac{3\bar m^{[3]}}{\hat m \bar m^2}L_6^r + 36 \frac{\bar m}{\hat m} L_7^r + \left(12\frac{\bar m}{\hat m}-1\right)L_8^r   \right] \nonumber\\
& + \left[ \frac{3\bar m^{[3]}}{m_u^3}\left(1-\frac{m_u}{m_d}\right)^2\left(1+\frac{m_u}{m_d}\right)  + \frac{4\bar m}{m_s}-3\right]\left( \mu_{\pi^0}^{}+2\mu_{\pi^+}^{} \right) \nonumber\\
& + \left[\frac{6\bar m^{[3]}}{m_s^3}\left(1-\frac{m_s}{m_d}\right)^2\left(1+\frac{m_s}{m_d}\right)+\frac{8\bar m}{m_u}-7 \right]\mu_{K^0}^{} \nonumber\\
& +  \left[\frac{6\bar m^{[3]}}{m_s^3}\left(1-\frac{m_s}{m_u}\right)^2\left(1+\frac{m_s}{m_u}\right)+\frac{8\bar m}{m_d}-7 \right]\mu_{K^+}^{} \nonumber\\
&+ \left[ \frac {3\bar{m}^{[3]} }{m_s^3}-\frac {4\bar{m}}{m_s} - \frac{\bar{m}^{[3]}(m_s +3\bar{m})^2}{m_s^2\bar{m}^2 (m_u + m_d+ 4m_s)}\right] \mu_{\eta}^{}
+\mathcal{O}\left(\frac{\delta^2}{m_s^2}\right)
\Bigg\} \nonumber\\
&+ \frac{3\bar m^4}{32\pi^2f_a^4} \Bigg[ \frac{3M_{\pi^+}^4}{m_u^2m_d^2} + \frac{2M_{K^+}^4}{m_u^2m_s^2} + \frac{2M_{K^0}^4}{m_d^2m_s^2}
+ \frac{\left(2m_uM_{K^0}^2 + 2m_dM_{K^+}^2 - m_sM_{\pi^+}^2\right)^2}{9m_u^2m_d^2m_s^2}  \Bigg].
\label{eq:lam}
\end{align}

The numerical evaluation requires the values of the quark mass ratios and of the LECs, which have been determined by the lattice QCD calculations and experimental data. A review of the present knowledge of the LECs appearing in the chiral Lagrangian for the meson sector can be found in Ref.~\cite{14Bijnens.Ecker149-174ARNPS}. Using the input values listed in Table~\ref{tab:pama}, we find the axion mass and the quartic axion self-coupling at NLO to be
\begin{eqnarray}\label{maNLO0}
m_a&=& 5.89(10) \mu\text{eV}
\cdot\frac{10^{12}~\text{GeV}}{f_a}~,\\
\lambda_4&=&
-\left(5.86(19)\cdot\frac{10^{-2}~\mathrm{GeV}}{f_a}\right)^4~, \label{lam0}
\end{eqnarray}
respectively.
Here we have used the charged pion mass in Eq.~\eqref{eq:mpifpi} for eliminating the overall $B_0(m_u +m_d)$ factor in $m_a^2$ and $\lambda_4$.
Although the difference between the charged and neutral pions from QCD is of $\order{\delta^2}$, the charged pion receives an electromagnetic contribution at LO. Such an effect to the quantities of interest here can be eliminated if using the neutral pion mass instead, which amounts to replacing $M_{\pi^+}^2$ by $M_{\pi^0}^2$ in Eqs.~\eqref{eq:masq} and \eqref{eq:lam} and adding the following terms inside the curly brackets of these two expressions~\cite{85Gasser.Leutwyler465-516NPB}:
\begin{equation}
    \frac{\left(M_{K^+}^2 - M_{K^0}^2\right)_\text{QCD}^2}{3 M_{\pi^0}^2 \left(M_\eta^2 - M_{\pi^0}^2 \right)} \left[
    1 + \frac{8}{3}\Delta_\text{GMO} + \frac{M_{K^0}^2}{8\pi^2 F_\pi^2} \left( 1 + 6\ln \frac{M_{K^0}^2}{M_\eta^2} \right) + \order{\hat m, m_s}
    \right],
\end{equation}
with
\begin{align}
    \left(M_{K^+}^2 - M_{K^0}^2\right)_\text{QCD} & = M_{K^0}^2 - M_{K^+}^2 - M_{\pi^0}^2 + M_{\pi^+}^2 \,, \nonumber\\
    \Delta_\text{GMO} &= \frac{2M_{K^0}^2 + 2M_{K^+}^2 - 2M_{\pi^+}^2 + M_{\pi^0}^2 -3 M_{\eta}^2}{ M_\eta^2 - M_{\pi^0}^2 },
\end{align}
where the electromagnetic effects have been taken into account.
As a result, the values in Eqs.~\eqref{maNLO0} and \eqref{lam0} become
\begin{eqnarray}\label{maNLO}
m_a&=& 5.71(9) \mu\text{eV}
\cdot\frac{10^{12}~\text{GeV}}{f_a}~,\\
\lambda_4&=&
-\left(5.77(18)\cdot\frac{10^{-2}~\mathrm{GeV}}{f_a}\right)^4~,
\end{eqnarray}
which are regarded as our results for these quantities and will be used in the following.

As we mentioned earlier, both the axion mass and its self-coupling are tightly related to the cumulants of the QCD topological charge distribution through the $\theta$-vacuum energy density, see Eq.~(\ref{malachic4}).
Thus, from Eq.~(\ref{eq:cumulants}) or (\ref{malachic4})
we can further extract the numerical values of the topological susceptibility $\chi_t$ and the normalized fourth cumulant $b_2=c_4/(12\chi_t)$~\cite{12Bernard.Descotes-Genon.ea80-80J} with the inclusion of isospin breaking effects at zero temperature, i.e.,
\begin{eqnarray}
\chi_t^{1/4}&=&\sqrt{m_af_a}=
75.6(6)~\mathrm{MeV}~,\\
 b_2&=&
\frac{\lambda_4f_a^2}{12m_a^2}=-0.028(3)~.
\end{eqnarray}
Since the masses of the octet of pseudoscalar mesons are well-known from experiments, the uncertainties
are in fact  dominated by the renormalized LEC $L_7^r$, while the subdominant  uncertainties
are from the  quark  mass  ratio $z=m_u/m_d$ and the LECs $L_6^r$ and $L_8^r$.
In comparison, the values of these quantities obtained here remain almost the same as the one in SU(2) case numerically, which are $\chi_t^{1/4} = 75.5(5)$~MeV and $b_2 = -0.029(2)$ ~\cite{16GrillidiCortona.Hardy.ea34-34J}.
And the result for the topological susceptibility is in perfect agreement
with recent $N_f=2+1+1$ lattice QCD simulation at the physical point giving
$\chi_t^{1/4}=75.6(1.8)(0.9)~\textrm{MeV}$~\cite{16Borsanyi.others69-71N}.
This indicates that the explicit inclusion of the strange quark degree of freedom does not induce large differences on the axion properties.
There are at least two compelling reasons accounting for this feature. First, the effects from  the heavier quark flavors have been largely included in the corresponding SU(2) LECs.
Second, in Ref.~\cite{16GrillidiCortona.Hardy.ea34-34J} the authors performed their numerical calculations with a  matching between two-flavor and three-flavor CHPT LECs.
Thus, the inclusion of the strange-quark degree of freedom does not change the results sizeably.
Yet, the expressions given here should be useful for chiral extrapolation of lattice results performed at unphysical quark masses, in particular when the up and down quark masses are close to the strange quark one.

\begin{table}
	\begin{centering}
\setlength\tabcolsep{4pt}
\renewcommand{\arraystretch}{1.4}
		\begin{tabular}{|c|c|c|c|c|c|c|}
			\hline
			  $z$ & $r$  & $M_{\pi^{+}}$ & $M_{\pi^0}$ & $M_{K^{+}}$  & $M_{K^{0}}$  & $M_\eta$ \tabularnewline
			\hline
			0.485(19) & 27.42(12) & 139.57 & 134.98 & 493.68(2)  & 497.61(1) & 547.86(2) \tabularnewline
			\hline
			 $F_\pi$ & $L_6^r$ &  $L_7^r$ & $L_8^r$ &  $C_7^W$ & $C_8^W$ &   \tabularnewline
			\hline
			92.28(9)  & 0.0(4) & $-0.3(2)$ & 0.5(2) & $\approx0$ & $0.60\pm0.20$ &   \tabularnewline
			\hline
		\end{tabular}
		\par\end{centering}
	\caption{\label{tab:pama} Numerical inputs used in this paper. The pion decay constant $F_\pi$, and experimental meson masses $M_P$ are in units of MeV, and are taken from Ref.~\cite{Tanabashi:2018oca}. The renormalized LECs $L_i^r$ are in units of $10^{-3}$; they correspond to values at  scale $\mu=770$~MeV and are taken from Ref.~\cite{14Bijnens.Ecker149-174ARNPS}. The NNLO anomalous LECs $C_7^W$ and $C_8^W$ are given in units of $10^{-3}$\,GeV$^{-2}$; for their determinations, see the text. For the quark mass ratios defined as $z =m_u/m_d$ and $r=m_s/\hat m$, we take the FLAG average of the $N_f=2+1$ lattice results~\cite{Aoki:2019cca}.  }
\end{table}

\section{Axion-photon coupling}
\label{garr-su3}

The axion-photon coupling is defined by the following Lagrangian (see, e.g., Refs.~\cite{85Kaplan215-226NPB,16GrillidiCortona.Hardy.ea34-34J,19Alonso-alvarez.Gavela.ea223-223EPJC}),
\begin{eqnarray}\label{garr-coupling}
\mathcal{L}_{a\gamma\gamma}= \frac{1}{4}g_{a\gamma\gamma} aF^{\mu\nu}\tilde{F}_{\mu\nu},
\end{eqnarray}
where $\tilde{F}_{\mu\nu}=\frac{1}{2}\varepsilon_{\mu\nu\rho\sigma}F^{\rho\sigma}$, with $F^{\mu\nu}$ the electromagnetic field tensor with the sign convention $\epsilon_{0123}=+1$.
Specifically, the axion-photon coupling is  given by
\begin{align}\label{garrOp4}
g_{a\gamma\gamma} &= \frac{\alpha_\text{em}}{2\pi f_a}\frac{\mathcal{E}}{\mathcal{C}} + g_{a\gamma\gamma}^\text{QCD}, \nonumber\\
g_{a\gamma\gamma}^\text{QCD} &= - \frac{\alpha_\text{em}}{2\pi f_a}  6 \langle\mathcal{X}_a Q^2\rangle + g_{a\gamma\gamma}^\text{mix}   = - \frac{\alpha_\text{em}}{2\pi f_a}\bigg(  \frac23 + 2 \mathcal{X}_u \bigg) + g_{a\gamma\gamma}^\text{mix},
\end{align}
where $\mathcal{E}/\mathcal{C}$ is the ratio of the electromagnetic and color anomaly coefficients, which is given by $\sum_n (Q_\text{PQ}Q^2)/\sum_n (Q_\text{PQ}T^2)$, with the sums running over all fermions with PQ charges $Q_\text{PQ}$, and $T^a$ the QCD color generators satisfying $\langle T^a T^b\rangle =T^2\delta^{ab}/2$. The value of $\mathcal{E}/\mathcal{C}$ depends on the specific axion models. The first term in $g_{a\gamma\gamma}^\text{QCD}$ is the contribution from the axial rotation of the quark fields, $q\to \exp\left(i\frac{a}{2f_a}\mathcal{X}_a\gamma_5 \right)q$ with $\langle \mathcal{X}_a\rangle=1$ (here we use the convention $\gamma^5=i\gamma^0\gamma^1\gamma^2\gamma^3$), which was introduced to eliminate the term $\frac{a}{f_a}\frac{\alpha_s}{8\pi} G_{\mu\nu}^c\tilde G^{c,\mu\nu}$ from the axion Lagrangian. The second term in $g_{a\gamma\gamma}^\text{QCD}$, $g_{a\gamma\gamma}^\text{mix}$, is the contribution from the $a$-$\pi^0$ and $a$-$\eta$ mixings, with the $\pi^0$ and $\eta$ coupled to two photons.

As discussed  below Eq.~\eqref{eq:dashen-nuyts}, there is a freedom of choosing the diagonal matrix $\mathcal{X}_a$ satisfying $\langle \mathcal{X}_a\rangle=1$. If it is chosen as $\mathcal{X}_a = \text{diag}\left\{\bar m/m_u, \bar m/m_d, \bar m/m_s \right\} =\bar m \mathcal{M}_q^{-1}$ as in Refs.~\cite{86Georgi.Kaplan.ea73-78PLB,87Kim1-177PR}, then $U=\tilde U=e^{i\Phi/F_0}$, see Eq.~\eqref{eq:Xf}, and there is no $a$-$\pi^0$ or $a$-$\eta$ mixing term in the LO chiral Lagrangian. One obtains the $\order{p^4}$ contribution to the model-independent $a\gamma\gamma$ coupling to be
\begin{equation}
    g_{a\gamma\gamma}^\text{QCD,(4)} = -\frac{\alpha_\text{em}}{2\pi f_a} \frac{2\left(m_u+3\bar m\right)}{3m_u}.
    \label{garrWZWop4}
\end{equation}
This result recovers the one derived in SU(2) CHPT~\cite{16GrillidiCortona.Hardy.ea34-34J} at $\mathcal{O}(p^4)$ in the limit of $m_s\to \infty$.

The same result can also be obtained by using other choices of $\mathcal{X}_a$. In that case,  one needs to consider  $a$-meson mixing.
The Wess--Zumino--Witten (WZW)  Lagrangian~\cite{71Wess.Zumino95-97PLB,83Witten422-432NPB} with an external photon field can be used to get the mixing contribution. The Lagrangian is given by~\cite{Kaymakcalan:1983qq,Meissner:1987ge,12Scherer.Schindler1-338LNP}
\begin{eqnarray} \label{12SchererLwzw}
\mathcal{L}_{\mathrm{WZW}}^\text{em} &=&
 -\frac{eN_c}{48\pi^2}\varepsilon^{\mu\nu\rho\sigma}A_\mu
 \left\langle Q\partial_\nu UU^{\dagger}\partial_\rho UU^{\dagger}\partial_\sigma UU^{\dagger}  +QU^{\dagger}\partial_\nu UU^{\dagger}\partial_\rho UU^{\dagger}\partial_\sigma U\right\rangle \nonumber\\
 && + i\frac{e^2N_c}{48\pi^2}\varepsilon^{\mu\nu\rho\sigma}  \partial_\nu A_\rho A_\sigma \left\langle 2Q^2(U\partial_\mu U^{\dagger}-U^{\dagger}\partial_\mu U)  -QU^{\dagger}Q\partial_\mu U+QUQ\partial_\mu U^{\dagger}\right\rangle,~~~~~
\end{eqnarray}
where $e>0$ is the electric charge unit, $Q$ and $N_c$ denote the usual diagonal quark charge matrix, $Q=\mathrm{diag}\{2/3,-1/3,-1/3\}$ for the three-flavor case, and the number of quark colors, respectively. Here the convention is such that $U$ transforms under SU(3)$_L\times$SU(3)$_R$ as $U\to g_R^{} Ug_L^\dag$ with $g_L$ and $g_R$ elements in SU(3)$_L$ and SU(3)$_R$, respectively.
According to Weinberg's power counting scheme, the above WZW Lagrangian starts to contribute from $\mathcal{O}(p^4)$.
The axion-meson mixing contribution can be obtained by substituting $U$ in the above Lagrangian by $\exp\left( -i\mathcal{Y}_a \frac{a}{f_a} \right)$ with $\mathcal{Y}_a = \mathcal{X}_a - \bar m \mathcal{M}_q^{-1}$.
One finds
\begin{eqnarray}\label{garrWZW}
g_{a\gamma\gamma}^{\text{mix}}=\frac{\alpha_{\text{em}}}{2\pi f_a}\bigg(2\mathcal{X}_u-2\frac{\bar{m}}{m_u}\bigg).
\end{eqnarray}
Using Eq.~\eqref{garrOp4}, one again gets the expression given in Eq.~(\ref{garrWZWop4}).

Our goal in this section is to compute the axion-photon coupling to $\mathcal{O}(p^6)$.
The chiral Lagrangian with a minimal set of terms in the anomalous-parity strong sector at $\mathcal{O}(p^6)$ has been given in Ref.~\cite{02Bijnens.Girlanda.ea539-544EPJC}, not only for SU(2) but also for SU($N$) with $N\geq3$. Based on the anomalous Lagrangians, several works have been done in the anomalous-parity sector~\cite{09Kampf.Moussallam76005-76005PRD,10Bijnens.Kampf220-223NPBS}.  In this work, only the terms proportional to $C_7^W$ and $C_8^W$ are relevant to the axion-photon coupling, which read
\begin{eqnarray}
\mathcal{L}_{\mathrm{ano}}^{(6)} = iC_7^W\varepsilon^{\mu\nu\rho\sigma}\left\langle \chi_{-}f_{+\mu\nu}f_{+\rho\sigma}\right\rangle +iC_8^W\varepsilon^{\mu\nu\rho\sigma}\left\langle\chi_{-}\right\rangle \left\langle f_{+\mu\nu}f_{+\rho\sigma}\right\rangle,
\end{eqnarray}
where $C_7^W$ and $C_8^W$ are two LECs. We have taken the same notation as in Ref.~\cite{02Bijnens.Girlanda.ea539-544EPJC}.

\begin{figure}
  \includegraphics[width=1.\textwidth]{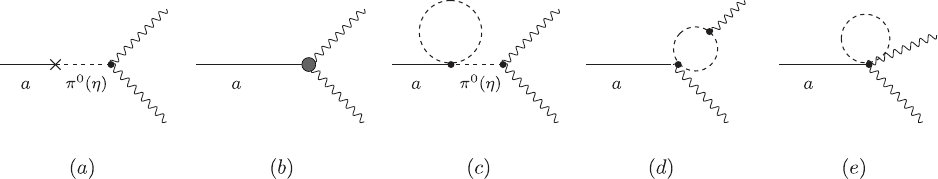}\\
  \caption{Feynman diagrams for the computation of the axion-photon coupling up to $\order{p^6}$. Here the dashed lines denote Goldstone bosons and wavy lines photons. Only pseudoscalar mesons are running in the loops. }\label{garrCoupling}
\end{figure}

In the following, we choose $\mathcal{X}_f = \bar m/m_f$ and $U=\tilde U$ for the computation of the $a\gamma\gamma$ coupling. With this convention the diagrams relevant for the computation of the $\mathcal{O}(p^6)$ corrections to $g_{a\gamma\gamma}$ are  depicted in Fig.~\ref{garrCoupling}:
(a) the axion-pion and axion-eta mass mixing from the NLO tree-level Lagrangian;
(b) the tree-level diagram from $\mathcal{L}_{\mathrm{ano}}^{(6)}$;
(c) one-loop diagrams with one vertex taken from $\mathcal{L}_{\mathrm{WZW}}$ and the other one taken from the LO chiral Lagrangian; the contributions from diagrams (d) and (e) exactly cancel with each other with the upper photon line in diagram (d) being on shell.
It is interesting to note that for the anomalous processes such as $\pi^0$, $\eta$ and $\eta'$ decaying into two photons, the one-loop contributions vanish when the up-down quark mass difference is neglected~\cite{85Donoghue.Holstein.ea2766-2769PRL,88Bijnens.Bramon.ea1453-1453PRL}.
Likewise, in the SU(2) case the sum of all one loop corrections
vanishes when both photons in the final state are on-shell~\cite{16GrillidiCortona.Hardy.ea34-34J}. However, in the SU(3) case, diagram (c) does contribute to the axion-photon coupling at $\mathcal{O}(p^6)$ when taking isospin breaking effects into account.
Note that the pion-eta mixing needs to be considered in order to keep $g_{a\gamma\gamma}$ scale-independent and UV finite.

Putting together all the pieces, we obtain the axion-photon coupling keeping all orders in strong isospin breaking up to $\mathcal{O}(p^6)$ as
\begin{eqnarray}
g_{a\gamma\gamma}
&=&\frac{\alpha_{\text{em}}}{2\pi f_a}\Bigg\{\frac{\mathcal{E}}{\mathcal{C}}
-\frac{2}{3}\frac{m_u+3\bar{m}}{m_u} \nonumber\\
&&
+\frac{1024\pi^2 }{3\hat m}\bar{m}M_{\pi^{0}}^2\left(C_7^W+3C_8^W\right)
-\frac{2\bar{m}M_{\pi^0}^2}{3\hat m}\left[\frac{f_{+}(\cos,\sin)}{\sqrt{3}M_{\eta}^2}+\frac{f_{-}(\sin,\cos)}{M_{\pi^0}^2} \right]
\Bigg\},~~~~\label{eq:garrorderp6}
\end{eqnarray}
where
\begin{eqnarray}
f_{\pm}(\cos,\sin)&=&\sqrt{3}\left(\mu_{\pi^0}^{}-\mu_\eta^{}\right)\cos(3\epsilon)\pm 3\left(\mu_{K^0}^{}-\mu_{K^+}^{}\right)\sin\epsilon -\sqrt{3}\left(\mu_{K^0}^{}+\mu_{K^+}^{}-2\mu_{\pi^+}^{}\right)\cos\epsilon\nonumber\\
&&-\frac{8M_{\pi^0}^2}{F_\pi^2\hat m}\left(3L_7^r+L_8^r\right) \left[2\sqrt{3}(\hat m-m_s)\cos\epsilon \pm 3\delta \sin\epsilon\right].~~~~
\end{eqnarray}
The functions $f_{\pm}(\sin,\cos)$ are equivalent to $f_{\pm}(\cos,\sin)$ with the sine and the cosine interchanged, i.e., $f_{\pm}(\sin,\cos)=f_{\pm}(\cos\rightarrow\sin,\sin\rightarrow \cos )$, and $\epsilon$ is the LO pion-eta mixing angle in the vacuum as can be obtained by setting $\theta=0$ in the expression of $\epsilon_\theta$ in Eq.~(\ref{eeTheta}).

For the parameters $C_7^W$ and $C_8^W$, it was argued in Ref.~\cite{09Kampf.Moussallam76005-76005PRD} that $C_7^W$ is largely suppressed compared to $C_8^W$ as the latter receives a strong contribution from the $\eta'$ while the former does not. The authors also suggested $|C_7^W|<0.1|C_8^W|$. We use $\Gamma(\eta\to\gamma\gamma)=\frac{M_\eta^3}{64\pi}|T_\eta|^2$ with the $\eta\to\gamma\gamma$ amplitude given by~\cite{09Kampf.Moussallam76005-76005PRD}
\begin{equation}
T_\eta = \frac{e^2}{\sqrt{3}F_\pi} \bigg[ \frac{F_\pi}{4\pi^2 F_\eta}(1+x_\eta) - \frac{64}{3} M_\pi^2 C_7^W + \frac{256}{3}(r-1) M_\pi^2 \left(\frac16 C_7^W + C_8^W\right) + \mathcal{O}(m_s^2) \bigg]
\end{equation}
to extract the value of $C_8^W$ from the measured value of the $\eta\to\gamma\gamma$ width: $(0.516\pm0.020)$~keV~\cite{Tanabashi:2018oca}. Following Ref.~\cite{09Kampf.Moussallam76005-76005PRD}, we take $F_\eta=(118.4\pm8.0)$~MeV and assign a 30\% uncertainty for the $\order{m_s^2}$ contribution compared to that of $\order{m_s}$, we get $C_8^W=(0.60\pm0.20)\times10^{-3}$~GeV$^{-2}$, as listed in Table~\ref{tab:pama}. We have set $C_7^W$ to 0 as its effect can be absorbed into the uncertainty of $C_8^W$.

With the input parameters presented in Table~\ref{tab:pama}, one gets
\begin{eqnarray} \label{garrNumer}
g_{a\gamma\gamma}=\frac{\alpha_{\text{em}}}{2\pi f_a}\bigg(\frac{\mathcal{E}}{\mathcal{C}}-1.94(3)\bigg)
=\left[ 0.203(3)\frac{\mathcal{E}}{\mathcal{C}}-0.394(7)\right] \frac{m_a}{\text{GeV}^2}.
\end{eqnarray}
The error for the axion-photon coupling is also dominated by the uncertainties of $C_8^W$, $z$ and $L_7^r$, which are of similar size.
From Eq.~(\ref{garrNumer}), we obtain $g_{a\gamma\gamma}\simeq1.5\times10^{-16}\ldots 1.5\times10^{-13}~\text{GeV}^{-1}$  for  the axion mass in the range  $1\ldots 1000~\mu\text{eV}$ with $\mathcal{E}/\mathcal{C}=8/3$. Especially for $m_a=6.7~\mu\text{eV}$, this equation predicts  $g_{a\gamma\gamma}\simeq1.0\times10^{-15}~\text{GeV}^{-1}$ for models with $\mathcal{E}/\mathcal{C}=8/3$ like the Dine--Fischler--Srednicki--Zhitnitsky (DFSZ) model~\cite{Shifman-1979if}, which is still in the allowed region by the recent axion dark matter search, with $m_a$ around 6.7~$\mu\text{eV}$~\cite{Lee-2020cfj}.

\begin{table}
	\begin{centering}
\setlength\tabcolsep{2pt}
\renewcommand{\arraystretch}{1.4}
		\begin{tabular}{|l|c|c|c|c|c|}
			\hline
			$N$ & $m_a$\,$\big[\mu$eV$\cdot\frac{10^{12}\,\text{GeV}}{f_a}\big]$ & $(-\lambda_4)^{1/4}$ [$10^{-2}$GeV$/f_a$]  & $g_{a\gamma\gamma}^\text{QCD}$ $\big[\frac{\alpha_\text{em}}{2\pi f_a} \big]$ & $\chi_t^{1/4}$ [MeV] & $b_2$ \tabularnewline
			\hline
			2~\cite{16GrillidiCortona.Hardy.ea34-34J} & 5.70(7) & 5.79(10) & $-1.92(4)$ & 75.5(5) & $-0.029(2)$ \tabularnewline
			\hline
			3 & 5.71(9) & 5.77(18) & $-1.94(3)$ & 75.6(6) & $-0.028(3)$  \tabularnewline

			\hline
		\end{tabular}
		\par\end{centering}
	\caption{Summary of the main numerical results of the present work shown in the third line. For comparison we also show the results in the second line obtained in the framework of the SU(2) CHPT~\cite{16GrillidiCortona.Hardy.ea34-34J}.  For the axion-photon coupling $g_{a\gamma\gamma}$, only the model-independent part, denoted by $g_{a\gamma\gamma}^\text{QCD}$, is shown.
    \label{tab:compa} }
\end{table}

The Primakoff effect plays a key role in axion searches. For example, the working principle for an axion helioscope~\cite{Sikivie-1983ip,Anastassopoulos-2017ftl} is that axions produced in the core of the Sun are converted back into photons in a strong magnetic field.
Clearly, if the ratio $\mathcal{E}/\mathcal{C}=2$, which is quite a possibility as shown by Kaplan in Ref.~\cite{85Kaplan215-226NPB}, then the $g_{a\gamma\gamma}$ would be highly suppressed. The axion detection using the Primakoff effect, such as microwave
cavity experiments, 
or light shining through wall experiments (for a recent review, see Ref.~\cite{18Irastorza.Redondo89-159PPNP})
would thus be extremely difficult. Here, we present the reference values of $g_{a\gamma\gamma}$ for $\mathcal{E}/\mathcal{C}=2$ and $8/3$:
\begin{eqnarray}
g_{a\gamma\gamma}=
\begin{cases}
+0.07(4)\times10^{-3}/f_a ,&\mathcal{E}/\mathcal{C}=2,\cr
+0.85(4)\times10^{-3}/f_a,&\mathcal{E}/\mathcal{C}=8/3.
\end{cases}
\label{garNumber}
\end{eqnarray}

With the expressions of the axion mass and the axion-photon coupling, it is straightforward to estimate the axion lifetime, namely,
\begin{eqnarray}
\tau_{a\rightarrow\gamma\gamma}&=&\frac{64\pi}{g_{a\gamma\gamma}^2m_a^3}
=\frac{3.2\times 10^{54}~\mathrm{s}}{\big[\mathcal{E}/\mathcal{C}-1.94(3)\big]^2}\Big(\frac{\mu \mathrm{eV}}{m_a}\Big)^5.~~
\end{eqnarray}
As the axion lifetime is inversely proportional to $m_a^5$, the axion is more stable when its mass is smaller.
The axion lifetime is estimated as $\tau_{a\rightarrow\gamma\gamma}\gtrsim 10^{33}~\mathrm{s}$ if the lower limit
 $f_a\gtrsim 0.5\times10^{9}~\mathrm{GeV}$ is employed. Such a cosmologically stable particle is a well-motivated cold dark matter candidate~\cite{16Marsh1-79PR,13Kawasaki.Nakayama69-95ARNPS}.

\section{Summary}\label{SUMMARY}

In this paper, we have calculated the  QCD $\theta$-vacuum energy and in turn the axion potential up-to-and-including NLO
corrections in SU($N$)
CHPT.
Unlike the SU(2) case, no analytic solutions exist for SU($N$) with $N\geq 3$.
We work out for the first time a recursion relation for $\phi_f$, up to an arbitrary order in $\theta$. Then, as an extension of Ref.~\cite{15Guo.Meisner278-282PLB}, by expanding the one-loop effective generating functional around the free-field configuration in a $\theta$-vacuum, we have calculated the $\theta$-vacuum energy density up NLO, including the one-loop contribution, in SU($N$) CHPT with $N$ non-degenerate quark flavors. With the recursion relation for the $\phi_f$ angles, one can compute any-order cumulants of the QCD topological charge distribution as well as the axion mass and self-couplings.

Since the QCD axion potential as a function of $a/f_a$ takes the same form as the QCD $\theta$-vacuum energy as a function of $\theta$, we have also calculated the axion mass and self-coupling to NLO from the SU(3) $\theta$-vacuum energy density taking into account the strong isospin breaking effects.
With the determination of the LECs from experimental data and lattice simulations,
we have further evaluated the numerical values for axion mass and self-coupling up to NLO, which are similar to those obtained in the SU(2) case in Ref.~\cite{16GrillidiCortona.Hardy.ea34-34J}.

We also computed the axion-photon coupling up to $\mathcal{O}(p^6)$. Numerically, it is given by $g_{a\gamma\gamma}=\frac{\alpha_{\text{em}}}{2\pi f_a}[\mathcal{E}/\mathcal{C}-1.94(3)]$,
which implies that if $\mathcal{E}/\mathcal{C}= 2$, the axion-photon coupling would be extremely small. In this case the axion searches using $g_{a\gamma\gamma}$, such as light shining through a wall  or microwave
cavity experiments, would be very difficult. This might also have an important impact on the axion electrodynamics as well as the possible existence of boson stars,
in which the axion-photon coupling plays a crucial role.

\section*{Acknowledgments}

The authors thank N. R. Acharya, S. Gonz\`alez-Sol\'is, M.-J. Yan and  B.-S. Zou for useful discussions.
We are grateful to Hao-Yang Cheng, Xu Feng and Tian Lin for pointing out a sign inconsistency in the previous version of the axion-photon coupling.
F.-K.G. is grateful to the hospitality of the Helmholtz Institut f\"ur Strahlen- und Kernphysik where part of the work was done. M.-L.D. would like to thank the hospitality of the Institute of Theoretical Physics where part of the work was done.
This work is supported in part by the China Postdoctoral Science Foundation
(Grant No.~2017M620920), by NSFC and DFG through funds provided
to the Sino-German Collaborative Research Center ``Symmetries and the Emergence of Structure in QCD" (NSFC
Grant No.~11621131001, DFG Grant No.~TRR110), by NSFC (Grant
No.~11835015 and No.~~11947302), by the CAS Key Research Program of Frontier Sciences (Grant No.~QYZDB-SSW-SYS013), by the CAS Center for Excellence in Particle Physics
(CCEPP), and also by the Scientific Research Fund of Hunan Provincial Education Department (Grant No.~19C0772).
The work of U.-G.M. was also supported in part by the CAS President's International Fellowship Initiative (PIFI) (Grant No.~2018DM0034), and by the VolkswagenStiftung (Grant No. 93562).

\appendix

\section{Full solution of the vacuum angles for the \texorpdfstring{SU$\bm{(N)}$}{SU(N)} case}
\label{app:general}

Let us derive the expressions of the vacuum angles $\phi_f$, and thus the LO vacuum energy, to all orders of $\theta$ for SU($N$) here. The starting equations are the SU($N$) version of Eqs.~\eqref{eq:dashen-nuyts}
\begin{eqnarray}
\begin{cases}
m_f\sin\phi_f = \text{constant},\cr
\displaystyle\sum_{f=1}^N \phi_f = \theta .
\end{cases}
\label{eq:DS_sun}
\end{eqnarray}
We use the following expansions,
\begin{eqnarray}
  \sin \phi_f &=& \sum_{n=0}^{\infty} s_{2n+1} \phi_f^{2n+1},~~\text{with}~s_{2n+1}\equiv\frac{(-1)^n}{(2n+1)!}, \nonumber\\
  \phi_f &=& \sum_{m=0}^{\infty} C_{f,2m+1} \theta^{2m+1} .
  \label{eq:sin}
\end{eqnarray}
Once we solve all the coefficients $C_{f,2m+1}$, we then get the general solution of $\phi_f$. Let
\begin{equation}
    m_f\sin \phi_f = \sum_{n=0}^{\infty}\alpha_{2n+1}\theta^{2n+1},
    \label{eq:sin2}
\end{equation}
then Eqs.~\eqref{eq:DS_sun} are decomposed into equations for each odd order of $\theta$.

At $\order{\theta}$, one has
\begin{eqnarray}
m_f C_{f,1} = \alpha_1,~~\text{and}~~
\sum_{f=1}^N C_{f,1} = 1 .
\end{eqnarray}
Thus, one gets
\begin{equation}
C_{f,1} = \frac{\bar m}{m_f} \label{eq:Cf1}
\end{equation}
with $\bar m$ defined in Eq.~\eqref{eq:mbar}.

At $\order{\theta^3}$, one has
\begin{eqnarray}
C_{f,3} +s_3 C_{f,1}^3  = \frac{\alpha_3}{m_f},~~\text{and}~~
\sum_{f=1}^N C_{f,3} = 0 ,
\end{eqnarray}
the solution of which is
\begin{equation}
    C_{f,3} = s_3 \left( \frac{\bar m}{m_f} \sum_{i=1}^N C_{i,1}^3 - C_{f,1}^3 \right). \label{eq:Cf3}
\end{equation}
Combining Eqs.~\eqref{eq:Cf1} and \eqref{eq:Cf3}, one gets the known result in Eq.~\eqref{phiSolu}.

At $\order{\theta^5}$, one has
\begin{equation}
C_{f,5} + 3 s_3 C_{f,1}^2 C_{f,3} + s_5 C_{f,1}^5  = \frac{\alpha_5}{m_f},~~\text{and}~~
\sum_{f=1}^N C_{f,5} = 0 ,
\label{eq:alpha5}
\end{equation}
the solution of which is
\begin{eqnarray}
  C_{f,5} = 3 s_3\left(\frac{\bar m}{m_f} \sum_{i=1}^N C_{i,1}^2 C_{i,3} - C_{f,1}^2 C_{f,3} \right)  + s_5 \left( \frac{\bar m}{m_f} \sum_{i=1}^N C_{i,1}^5 - C_{f,1}^5 \right).
\end{eqnarray}
Notice that for the expansion of $\sin\phi_f$ in powers of $\theta$ in Eq.~\eqref{eq:sin2}, the terms at $\order{\theta^{2n+1}}$ are closely related to the partition of $2n+1$ into odd parts  (e.g., the partitions of 5 into odd parts include 5, $3+1+1$ and $1+1+1+1+1$, see the left side of Eq.~\eqref{eq:alpha5}) studied in number theory.

One can go to higher orders and solve for $C_{f,2n+1}$ in the same way.
Finally, one gets the recursion relation for all the coefficients as
\begin{eqnarray}
 C_{f,2n+1} = \sum_{t=1}^{n} \sum_{(k_1,\ldots,k_t)}  s_{K_t}^{}\binom{K_t}{k_1,\ldots,k_{t}} \Bigg[ \frac{\bar m}{m_f} \sum_{i=1}^N \prod_{j=1}^{t} C_{i,2j-1}^{k_j}
 - \prod_{j=1}^{t} C_{f,2j-1}^{k_j}  \Bigg],
\end{eqnarray}
where $k_j$ are nonnegative integers, $K_t \equiv\sum_{j=1}^t k_j$, $\sum_{(k_1,\ldots,k_t)}$ means that the sum runs over all possibilities of $k_j$ satisfying $k_1+\cdots+(2t-1)k_t=2n+1$, and $\binom{K_t}{k_1,\ldots,k_{t}}= K_t!/(k_1!\cdots k_t!)$ are the multinomial coefficients.



\end{document}